\documentclass[aps,prb,twocolumn,longbibliography]{revtex4-2}
\usepackage[utf8]{inputenc}

\usepackage[colorlinks=true,citecolor=blue]{hyperref}
\usepackage{graphicx}
\usepackage{physics}
\usepackage{bm}
\usepackage{amsmath}
\usepackage{comment}
\usepackage{amssymb}
\usepackage{qcircuit}
\DeclareGraphicsExtensions{.pdf,.jpg,.eps}
\usepackage{xcolor}

\newcommand{\beq}{\begin{eqnarray}}
\newcommand{\eeq}{\end{eqnarray}}

\begin{document}

\author{Marcel Niedermeier}
\affiliation{Department of Applied Physics, Aalto University, 02150 Espoo, Finland}

\author{Jose L. Lado}
\affiliation{Department of Applied Physics, Aalto University, 02150 Espoo, Finland}

\author{Christian Flindt}
\affiliation{Department of Applied Physics, Aalto University, 02150 Espoo, Finland}

\title{Simulating the quantum Fourier transform, Grover’s algorithm, and the quantum counting algorithm with limited entanglement using tensor-networks}

\begin{abstract}
Quantum algorithms reformulate computational problems as
quantum evolutions in a large Hilbert space.
Most quantum algorithms assume that the time-evolution is perfectly unitary and that the full Hilbert space is available. However, in practice, the available entanglement may be limited, leading to a reduced fidelity of the quantum algorithms. To simulate the execution of quantum algorithms with limited entanglement, tensor-network methods provide a useful framework, since they allow us to restrict the entanglement in a quantum circuit. Thus, we here use tensor-networks to analyze the fidelity of the quantum Fourier transform, Grover’s algorithm, and the quantum counting algorithm as the entanglement is reduced, and we map out the entanglement that is generated during the execution of each algorithm. In all three cases, we find that the algorithms can be executed with high fidelity even if the entanglement is somewhat reduced. Our results are promising for the execution of these algorithms on future quantum computers, and our simulation method based on tensor networks may also be applied to other quantum algorithms.
\end{abstract}
\date{\today}

\maketitle

\section{Introduction}

Noisy intermediate-scale quantum computers are limited by decoherence effects and loss of entanglement \cite{noise_ind_loss_entanglement, preservation_ent_local_noisy_channels, Guo2023}.
Error-corrected qubits \cite{Google_QEC, IBM_QEC, QEC_book} would provide a strategy to mitigate decoherence, yet the current capabilities have not yet reached the sizes required for the execution of quantum algorithms such as Shor's factoring algorithm \cite{Shor}, Grover's search algorithm \cite{Grover1996}, and the quantum Fourier transform~\cite{QFT_initial}. Given the limitations of current quantum hardware, an important practical question is how well those quantum algorithms, as well as others, can be executed on quantum computers, where the entanglement between qubits may be lost~\cite{NISQ_algorithms, NISQ_techniques, noise_frontier_quantum_sup, evidence_exp_adv_quantum_chemistry}.

\begin{figure}[t!]
   \centering
\includegraphics[width=0.98\columnwidth]{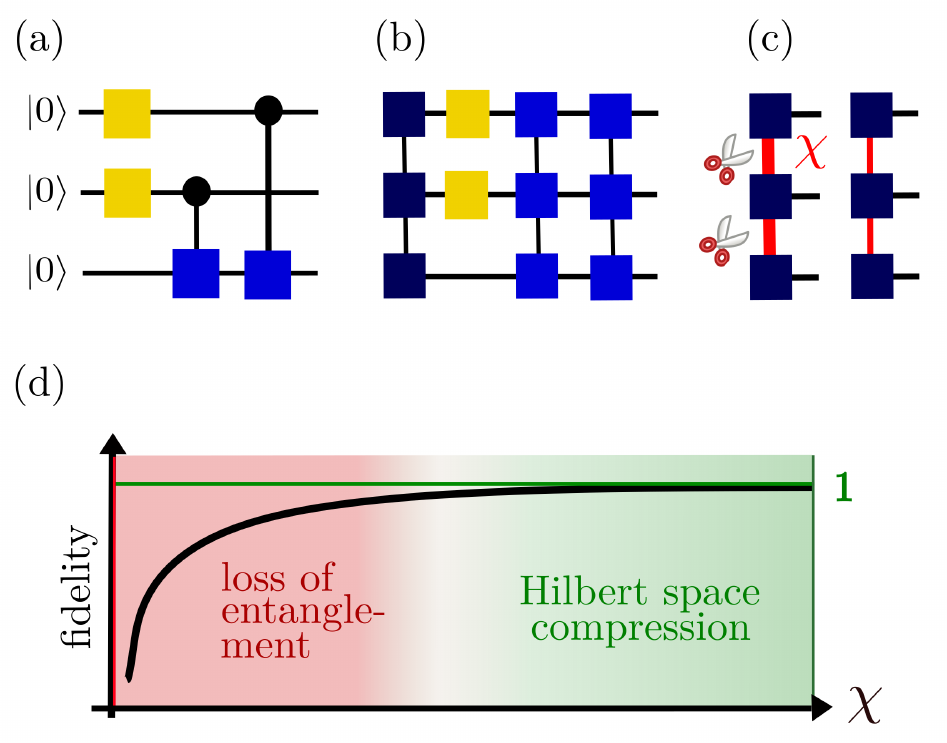}
   \caption{Tensor-network simulations of quantum algorithms with limited entanglement. (a) Example of a quantum circuit that can be simulated. (b) The quantum circuit is converted into an equivalent representation based on matrix product states. (c) The matrix product states are characterized by their bond dimension, $\chi$, which quantifies the amount of entanglement they can describe. (d)  In our simulations, we first reduce the bond dimension so that unnecessary parts of the Hilbert space are removed (green region). We then further reduce the bond dimension, so that relevant entangled states (red region) are removed, and entanglement is lost.}
   \label{fig:overview}
\end{figure}

There are several strategies for simulating the loss of entanglement in a {non-perfect} quantum computer. Tensor networks, including matrix product states~\cite{Baxter1968, fin_corr_states, Vidal_MPS, Schollwock2011, TN_nutshell, TN_developments}, have been shown to provide a suitable framework for the simulation of quantum circuits with limited entanglement~\cite{PRX_what_limits, DMRG_QC, Noisy_QC_MPDO, eff_class_sim_noisy_QC, TN_simulation_exa, validating_quantum_classical_programming, QC_simulation_class_supercomputer, Tensor_circuit, Sim_QC_TTN, QSim_big_batch, sampling_sycamore_TN, class_sim_quantum_supremacy, lossy_boson_sampling_MPO, efficient_2D_TN_sim_quantum_systems, Simulating_QC_MERA, heuristics_eff_par_TN_contr, Shor_alg_MPS, woolfe2017, Stoudenmire2023}. By controlling the bond dimension of the matrix product states, a limit on the entanglement can be enforced~\cite{PRX_what_limits}. In particular, tensor-network simulations of quantum circuits have found fidelities that are similar to those observed in experiments~\cite{PRX_what_limits}. In addition, connections between limited entanglement and classical simulations of quantum computers have been explored \cite{Shinaoka2022, adaptive_QC, eff_contr_QC_not_much_entanglement, simulation_QC, Markov2008, tensor_decomp_QC, sim_QC_eff_TN_algorithms}, and low-rank approximations of the quantum Fourier transform~\cite{QFT_small_entanglement} and Grover's algorithm~\cite{low_rank_approx_sim_QC, Stoudenmire2023} have put these ideas into practice. 

In this work, we use matrix product states to restrict the entanglement in the simulation of three quantum algorithms, allowing us to investigate their fidelity, performance, and entanglement propagation on quantum computers with limited entanglement. This idea is illustrated in Fig.~\ref{fig:overview}, showing how the quantum circuit in Fig.~\ref{fig:overview}a is represented by the matrix product states and operators in Fig.~\ref{fig:overview}b. By reducing the bond dimension of the matrix product states, Fig.~\ref{fig:overview}c, we can simulate the generic loss of entanglement and evaluate the fidelity of the quantum algorithms. In particular, as illustrated in Fig.~\ref{fig:overview}d, we first reduce the bond dimension, so that unexplored parts of the Hilbert space are removed (green region) without affecting the fidelities. This reduction of the Hilbert space provides a significant speed-up of our simulations and forms the basis of most calculations based on matrix product states. However, we then further reduce the bond dimension, which not only provides an additional speed-up, but also removes relevant entangled states (red region). Using this approach, we evaluate the fidelity of the quantum Fourier transform, Grover's search algorithm, and the quantum counting algorithm~\cite{Brassard1998} with limited entanglement. The quantum Fourier transform is an important building block for other quantum algorithms such as Shor's algorithm, while the quantum counting algorithm is a useful application of the more general quantum phase estimation algorithm \cite{Kitaev_QPE, nielsen_chuang_2010}. We also investigate the fidelity of the approximate quantum Fourier transform, where small qubit rotations in the exact quantum Fourier transform are omitted~\cite{AQFT_decoherence}. From a computational point of view, our simulations are efficient, since the quantum algorithms only explore a tiny fraction~($1/10^{4}$) of the  Hilbert space, allowing us to work with a low bond dimension.
For each algorithm, we map out the entanglement that is generated in the circuit, and we find that all three algorithms can be executed with high fidelity even at a moderate loss of entanglement.

The rest of our paper is organized as follows. In Sec.~\ref{MPSmethods}, we briefly review the theory of matrix product states and matrix product operators and explain how they can be used to simulate quantum algorithms. In Sec.~\ref{Simulation}, we simulate the execution of the quantum Fourier transform, Grover’s search algorithm, and the quantum counting algorithm, and we analyze their fidelity as a function of the bond dimension to mimic the loss of entanglement. Finally, in Sec.~\ref{conclusion}, we provide our conclusions together with an outlook on possible directions for the future. Some technical details are described in the Appendices.

\section{Tensor-network simulations}
\label{MPSmethods}

\begin{figure}
    \centering
    \includegraphics[width=0.98\columnwidth]{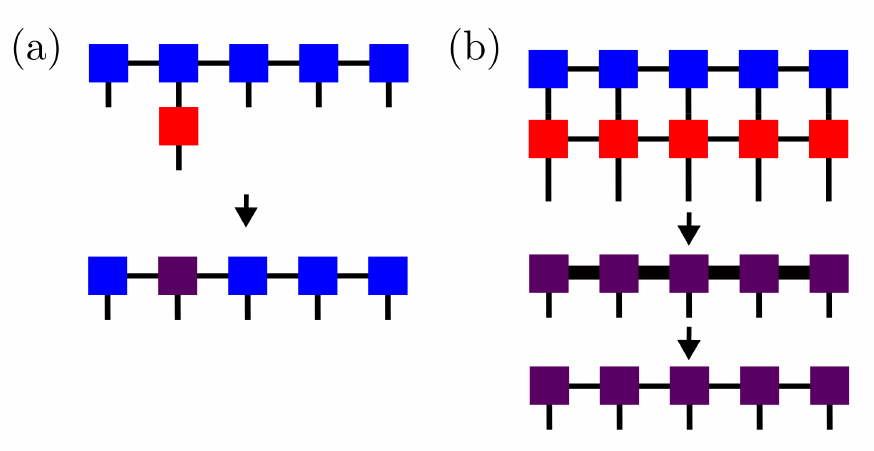}
    \caption{Single- and multi-qubit gates. (a) If a single-qubit gate is applied, it is sufficient only to update the tensor for the corresponding qubit. (b) The application of a multi-qubit gate leads to a higher bond dimension, which is then truncated to the largest allowed value. The value of the bond dimension is indicated by the thickness of the connecting lines.}
    \label{fig:MPS_MPO_application}
\end{figure}

\begin{figure*}
    \centering
    \includegraphics[width=0.98\textwidth]{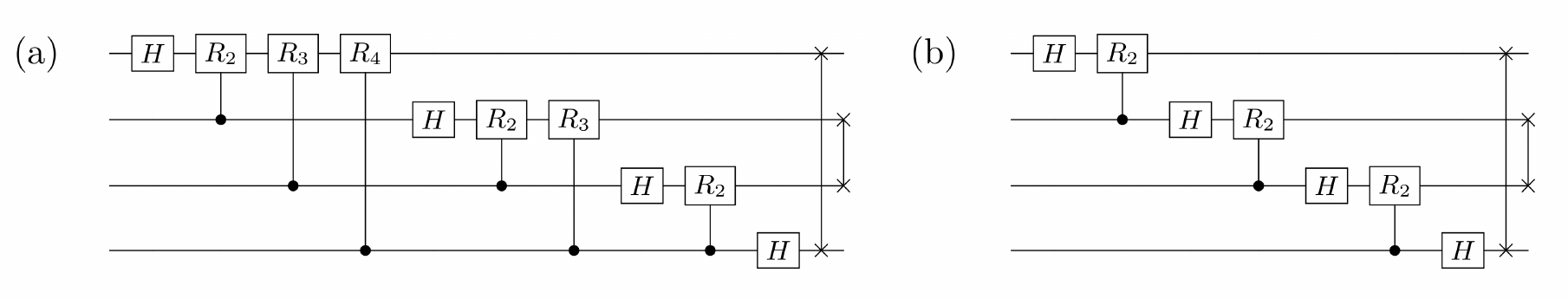}
    \caption{Quantum circuit for the quantum Fourier transform. (a) The quantum Fourier transform, here with $N=4$ qubits, is constructed from Hadamard gates, controlled rotations, and \texttt{SWAP} gates at the end. (b) In the approximate quantum Fourier transform, rotations below a certain angle are omitted so that only nearest-neighbor control is required in this example.}
    \label{fig:QFT_circuit}
\end{figure*}

We employ matrix product states to represent the state of a quantum computer as it executes a quantum algorithm~\cite{Vidal_MPS, Schollwock2011}. A generic quantum state can be expressed in the computational basis $\{|\boldsymbol{\sigma}\rangle= \ket{\sigma_1, \sigma_2, ..., \sigma_N}\}$, where $\sigma_i=0,1$ is the value of qubit number $i=1,\ldots,N$. We can then write the state of the quantum computer as
\begin{equation}
\ket{\Psi} = \sum_{\boldsymbol{\sigma}} C^{\boldsymbol{\sigma}}|\boldsymbol{\sigma}\rangle,
\end{equation}
where $C^{\boldsymbol{\sigma}}$ are the expansion coefficients. Importantly, the expansion coefficients $C^{\boldsymbol{\sigma}}$ can be decomposed as a chain of $N$ tensors $A_i$, reading
\begin{equation}
     C^{\boldsymbol{\sigma}} = \sum_{{\boldsymbol{\chi}}} [A_1]_{\chi_1}^{\sigma_1} [A_2]_{\chi_1 \chi_2}^{\sigma_2} ... [A_N]_{\chi_{N-1}}^{\sigma_N},
\end{equation}
where the vector $\boldsymbol{\chi}=(\chi_1, \chi_2, \ldots, \chi_{N-1})$ contains the bond dimensions of each tensor. This representation corresponds to a matrix product state with open boundary conditions and is always guaranteed to be exact if the bond dimensions $\chi_i$ are chosen exponentially large in the system size. In particular, the description of highly-entangled states requires large bond dimensions. For lower bond dimensions, with all bond dimensions bounded from above as $\chi_j \leq \chi_{\text{max}}$, highly entangled states are effectively excluded. We furthermore assume that the matrix product state $\ket{\Psi}$ is always normalized and brought into a suitable canonical form. This can be achieved by performing a sweep of successive singular value decompositions on each component tensor~$A_i$. 

In the following, we simulate the three quantum algorithms with different bond dimensions, corresponding to different upper bounds on the maximal entanglement entropy that can be reached. This approach will allow us to evaluate the fidelity of each algorithm if the entanglement is limited~\cite{PRX_what_limits}.
All of our calculations are performed with the \texttt{ITensors} package~\footnote{To use our publicly available code built on top of \texttt{ITensors}, please follow the instructions in Ref.~\onlinecite{QSim_github}.} and take less than a few hours of computing time on a single-core node \cite{ITensor, Itensors, QSim_github}. Before proceeding to our simulations, we first briefly discuss the implementation of single- and multi-qubit gates in the framework of matrix product states.

We first consider the application of a (non-entangling) single-qubit gate $M$ to a quantum state $\ket{\Psi}$. Since the gate only acts on a single tensor in the matrix product state, it is sufficient to update this tensor and replace the original tensor $A$ by the updated one $A'$  as
\begin{equation}
    [A_i]_{\chi_{i-1} \chi_i}^{\sigma_i} \rightarrow \sum_{\sigma_i} M^{\sigma_i' \sigma_i} [A_i]_{\chi_{i-1} \chi_i}^{\sigma_i} \equiv [A_i']_{\chi_{i-1} \chi_i}^{\sigma_i'}.
\end{equation} 
This procedure is illustrated in Fig.~\ref{fig:MPS_MPO_application}a. Multi-qubit gates, such as singly- and multiply-controlled gates, are constructed as matrix product operators. In analogy to matrix product states, a quantum operator
\begin{equation}
O = \sum_{\boldsymbol{\sigma}, \boldsymbol{\sigma'}} O^{\boldsymbol{\sigma},\boldsymbol{\sigma'}}  \!\ket{\boldsymbol{\sigma}} \!\bra{\boldsymbol{\sigma'}}
\end{equation}
can be parametrized in terms of a collection of tensors $W_i$, so that
\begin{equation} O^{\boldsymbol{\sigma},\boldsymbol{\sigma'}} = \sum_{\boldsymbol{\chi}} [W_1]_{\chi_1}^{\sigma_1, \sigma_1'} [W_2]_{\chi_1 \chi_2}^{\sigma_2, \sigma_2'} ... [W_N]_{\chi_{N-1}}^{\sigma_N, \sigma_N'}. 
\end{equation}
This construction ensures that the structure of the matrix product state is maintained if a matrix product operator is applied to it. In addition, using matrix product operators to represent multi-site gates allows us to implement arbitrary controlled gates in a unified manner which is independent of the separation between the control and the target qubit. Alternatively, it is possible to circumvent the matrix product operator construction by representing gates that act on neighboring qubits as rank-$4$ tensors~\cite{PRX_what_limits}. However, this approach requires the inclusion of \texttt{SWAP} gates to execute controlled operations, if the separation between control and target qubit is greater than one. As such, the method is more complicated for the representation of arbitrary controlled gates. Further details on the explicit construction of quantum gates in terms of matrix product operators are given in App.~\ref{appendix:appA}. 

In general, the application of a matrix product operator to a multiplicative matrix product state leads to an increased bond dimension \cite{Schollwock2011}. To maintain a given bond dimension, we therefore truncate the bond dimension after each application of an operator as shown in Fig.~\ref{fig:MPS_MPO_application}b. Here, when applying a multi-qubit gate, we employ a variational algorithm to find the best approximation of the quantum state for a given bond dimension~\cite{ Paeckel2019}. To perform a measurement on a quantum circuit represented by a matrix product state, we need to draw a sample from the probability distribution it represents. This can be achieved efficiently without performing a full contraction to recover the exact state vector~\cite{Ferris2012, Itensors}. We summarize this algorithm in App.~\ref{appendix:app_sampling}. Measurement statistics are obtained by repeating this procedure many times and constructing a histogram of the outputs.

\begin{figure*}
    \centering
    \includegraphics[width=0.98\textwidth]{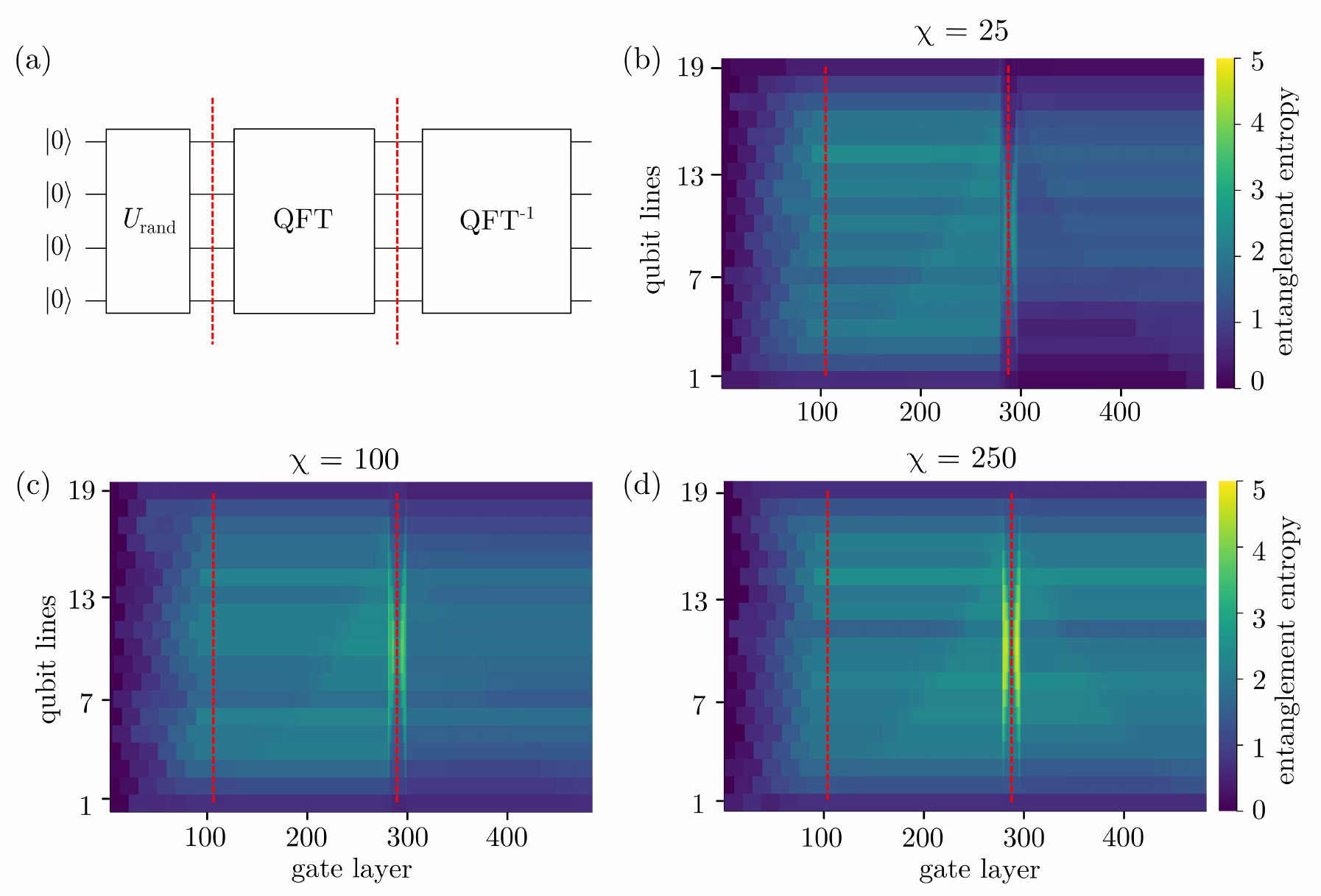}
    \caption{Entanglement in the quantum Fourier transform. (a) Quantum circuit for evaluating the fidelity of the quantum Fourier transform. The first gate, $U_\mathrm{rand}$, generates a random input state for the quantum Fourier transform (QFT), which is subsequently reversed. The fidelity is given in terms of the overlap between the random input state and the output of the circuit. (b-d) Evolution of the entanglement entropy throughout the circuit with increasing bond dimension, $\chi = 25, 100, 250$. 
    }
    \label{fig:QC_QFT}
\end{figure*}

\section{Quantum Algorithms}
\label{Simulation}

Below, we consider the quantum Fourier transform, Grover's search algorithm, and the quantum counting algorithm. In each case, we briefly review the algorithm before simulating its execution with limited entanglement.

\subsection{Quantum Fourier transform}

The quantum Fourier transform extends the classical discrete Fourier transform to the quantum realm. As its input, it takes a quantum state with $N$ qubits,
\begin{equation}
\ket{\Psi} = \sum_{\boldsymbol{\sigma}} C^{\boldsymbol{\sigma}}|\boldsymbol{\sigma}\rangle= \sum_{l=0}^{2^N-1} C_l \ket{l},
\end{equation}
where we have associated each state in the computational basis with an equivalent decimal number, $\{|\boldsymbol{\sigma}\rangle \} =  \{\ket{l}, | l = 0, ..., 2^N-1 \}$. The quantum Fourier transform of each basis state is now defined as
\begin{equation}
    \mathrm{QFT} \ket{l} = \frac{1}{\sqrt{2^N}} \sum_{m= 0}^{2^N-1}  e^{\frac{2\pi i}{2^N} l m} \ket{m},
\end{equation}
and the transformation of the general state becomes
\begin{equation}
    \mathrm{QFT} \ket{\Psi} = \frac{1}{\sqrt{2^N}} \sum_{l, m= 0}^{2^N-1}  C_l e^{\frac{2\pi i}{2^N} l m} \ket{m}.
\end{equation}
Figure~\ref{fig:QFT_circuit}a shows an example of the quantum Fourier transform with $N=4$ qubits. The circuit is constructed from controlled rotations of target qubits along the $z$-axis of the Bloch sphere as
\begin{equation}
    R_k = \begin{bmatrix}
           1 & 0 \\
           0 & e^{\frac{2\pi i}{2^k}}
     \end{bmatrix}
     \label{eq:Rk}
\end{equation}
together with \texttt{SWAP} gates at the end that reverse the order of the qubits. For large values of $k$, the rotations are small, and they can, to a good approximation, be neglected \cite{QFT_initial, AQFT_decoherence}. That procedure defines the approximate quantum Fourier transform, whose circuit is depicted in Fig.~\ref{fig:QFT_circuit}b, where controlled rotations with $k>2$ have been omitted. In the following,  we first simulate the execution of the quantum Fourier transform with limited entanglement before considering the approximate transformation.

For the classical Fourier transform, one is often interested in a one-to-one mapping between the time domain and the frequency domain, for example, to construct the entire frequency spectrum of an input signal. By contrast, in the context of quantum phase estimation and similar algorithms, the quantum Fourier transform is used to extract a single or a narrow range of frequencies.  Approximating a phase with a maximum error of $\epsilon $ requires about $-\lceil \log_2(\epsilon)\rceil$ qubits. For instance, a maximum error of $\epsilon =10^{-9}$ requires $30$ qubits, and we will therefore limit our simulations to this number of qubits.

It has been observed that the quantum Fourier transform does not rely on a high degree of entanglement~\cite{quantum_FFT_classical_simulation, Yoran2007,woolfe2017,QFT_small_entanglement}. The main exception is the use of \texttt{SWAP} gates at the end of the circuit to reorder the qubits. However, the action of the \texttt{SWAP} gates can also be obtained by redefining the circuit order by reversing the measurement sequence of the qubits, if the quantum Fourier transform is performed at the end of an algorithm.

In the present work, we consider the worst-case scenario, where the input state is prepared by a random quantum circuit of a given bond dimension, and is therefore highly unstructured. It can then be expected that for any application of the quantum Fourier transform to a structured input state, higher fidelities will be obtained. This expectation is analogous to classical signal processing: a random signal is complicated to decompose into frequencies, whereas a (structured) signal containing only a few distinct wavelengths will be easier to decompose into these. To prepare the random input state, we chose the same circuit as in Ref.~\cite{PRX_what_limits}, which models random quantum circuits used in previous works and experiments \cite{sycamore_quantum_supremacy, sampling_sycamore_TN}. It has been shown that the output distribution of such circuits quickly converges to the Porter-Thomas distribution after only a few gate layers \cite{PRX_what_limits}.

\begin{figure*}[t!]
\includegraphics[width=0.98\textwidth]{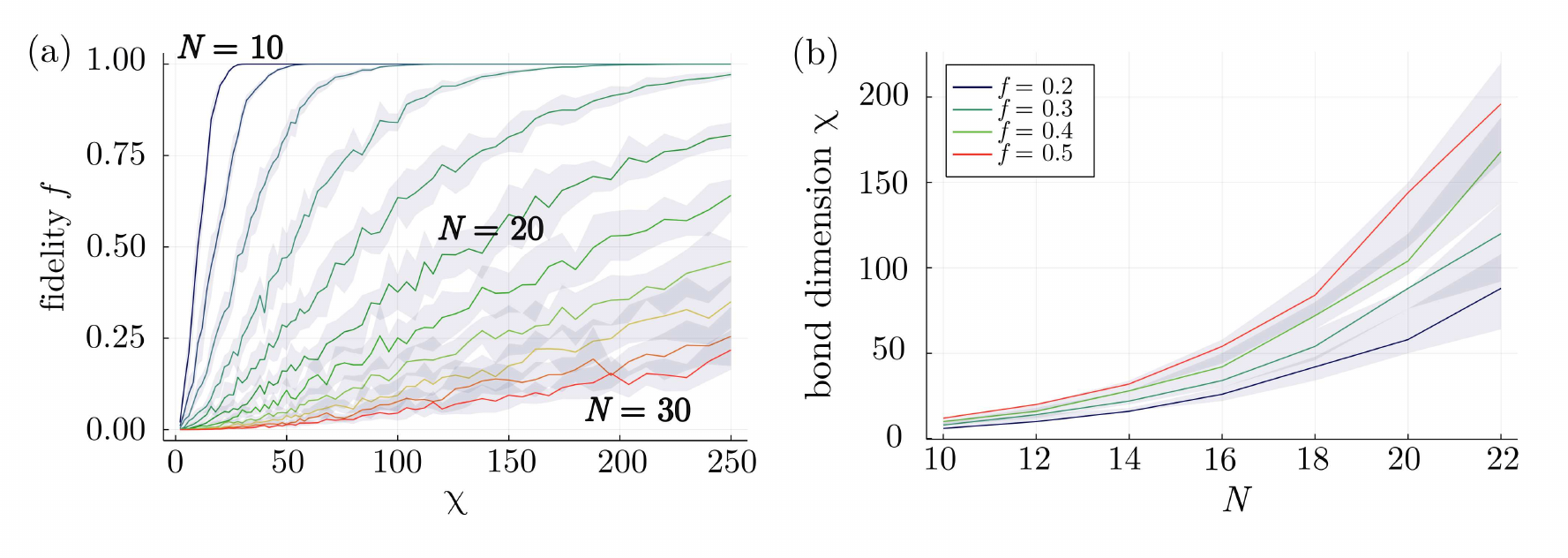}
      \label{fig:QFT_fid_2}
    \caption{Fidelity of the quantum Fourier transform. (a) The fidelity as a function of the bond dimension for different numbers of qubits. The results are averaged over $10$ random initial states with the standard deviations indicated by error ribbons. (b)~The required bond dimension to reach a given fidelity for different numbers of qubits. }
    \label{fig:QFT_fid}
\end{figure*}

We now define the fidelity as
\begin{equation}
    f = |\braket{\Psi_0'}{\Psi_0}|^2,
\end{equation}
where  $\ket{\Psi_0}$ is the random input state, and 
\begin{equation}
 \ket{\Psi_0'} =  \mathrm{QFT}^{-1}  \mathrm{QFT} \ket{\Psi_0}
\end{equation}
is the state after the quantum Fourier transform has been applied and reversed. In an ideal quantum computer, the fidelity would be one, whereas a reduction of the entanglement is expected to lower the fidelity. Figure~\ref{fig:QC_QFT}a shows the quantum circuit for obtaining $\ket{\Psi_0'}$ and finding the fidelity. We initialize the random circuit with $20$ layers of alternating single-qubit gates and \texttt{CNOT} gates~\cite{PRX_what_limits}. We keep track of the total number of multi-site gates, such that the generation of the random 20-qubit state in Fig.~\ref{fig:QC_QFT}a requires $105$ gate layers. We then average over 10 random states and evaluate the mean and the standard deviation of the fidelity.  

Before discussing the fidelity, we investigate how the entanglement entropy evolves throughout the quantum circuit. The entanglement entropy is defined as
\begin{equation}
    \mathcal S_L=-\mathrm{Tr}\{ \rho_L\ln\rho_L\},
\end{equation}
where $\rho_L$ is the reduced density of the first $L$ qubits in the circuit, obtained by tracing out the remaining $N-L$ qubits. The entanglement entropy vanishes for product states, while it has the maximum value of $\mathcal S^{\mathrm{max}}_L=L\ln{2}$ for highly entangled states of $L\leq N/2$ qubits. For $L>N/2$, the maximum entanglement entropy is $\mathcal S^{\mathrm{max}}_L=(N-L)\ln{2}$. In Figs.~\ref{fig:QC_QFT}b,c,d, we consider a circuit with $N=20$ qubits that we describe using three different bond dimensions. The vertical axis indicates the value of $L$ at which we partition the circuit. We can clearly distinguish three regions, indicated by dashed vertical lines, corresponding to the generation of the random state, the quantum Fourier transform, and its inverse. To begin with, the entanglement grows as an initial state is randomly generated for each panel. The entanglement continues to grow slowly as the quantum Fourier transform is applied, and it reaches a maximum around the \texttt{SWAP} gates, in accordance with the findings of Ref.~\cite{QFT_small_entanglement}. 

\begin{figure}
\includegraphics[width=0.98\columnwidth]{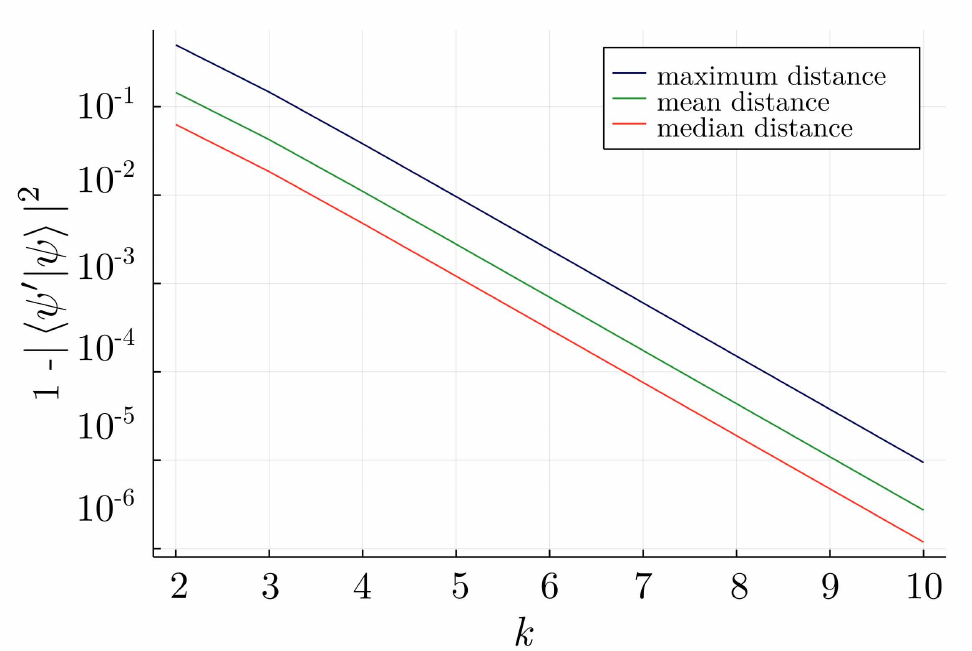}%
    \centering
    \caption{Controlled rotations. We show the distance between a rotated two-qubit state  $\ket{\Psi'}=CR_k\ket{\Psi}$ and the original state $\ket{\Psi}$ as a function of $k$, which controls the rotation angle. We show the maximum distance for all possible two-qubit states together with the mean and median values, see App~\ref{appB}.}
\label{fig:controlled_z_rotation_fidelity}
\end{figure}

\begin{figure*}
\includegraphics[width=0.98\textwidth]{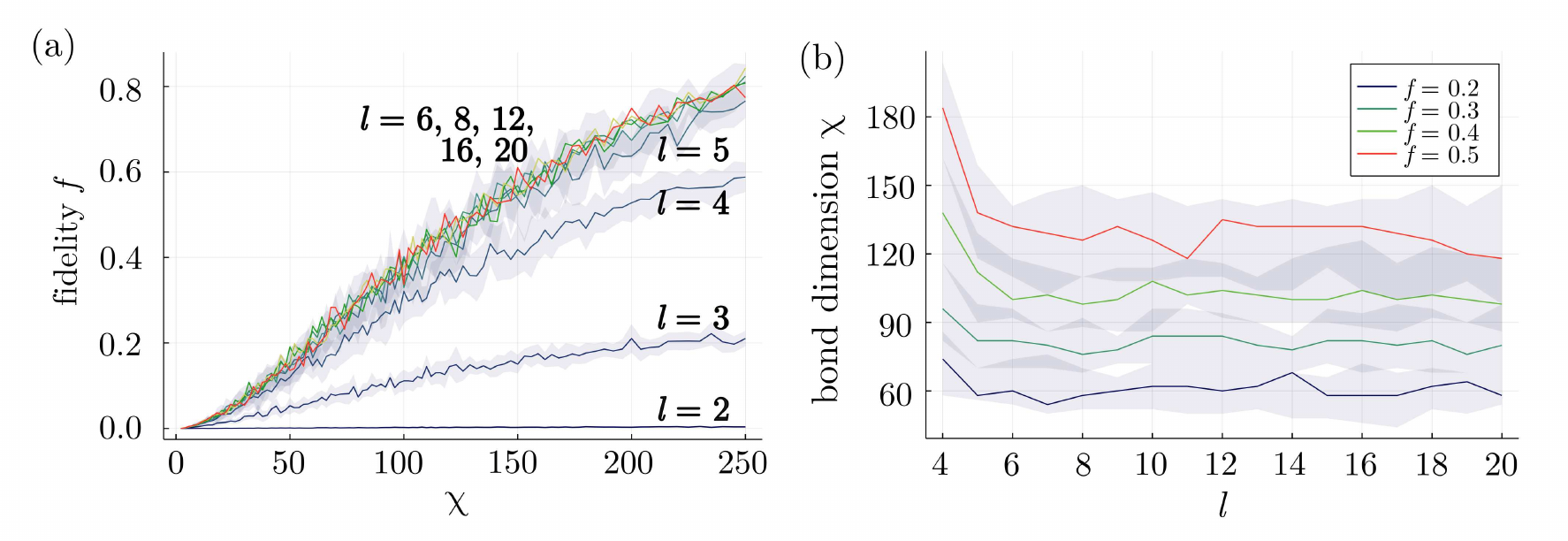}
    \caption{Fidelity of the approximate quantum Fourier transform. (a) The fidelity as a function of the bond dimension for $N=20$ qubits and different cut-offs $l$ of the rotations. The results are averaged over 10 random initial states with the standard deviations indicated by error ribbons. (b) The required bond dimension to reach a given fidelity as a function of the cut-off.}
    \label{fig:AQFT_fid}
\end{figure*}

Figure~\ref{fig:QFT_fid} displays the fidelity of the quantum Fourier transform. In Fig.~\ref{fig:QFT_fid}a, we show it as a function of the bond dimension for different numbers of qubits. With a small number of qubits, the fidelity quickly approaches one, indicating that the algorithm can be executed with a high fidelity even if the loss of entanglement is substantial. By contrast, with more qubits, a much higher bond dimension is required, implying that entangled states play an important role.  In  Fig.~\ref{fig:QFT_fid}b, we show the required bond dimension to reach a given fidelity as a function of the qubit number. The figure demonstrates that the required bond dimension grows faster than linearly with the number of qubits, and we find that our results are well-captured by the scaling $\chi\propto N^{3.5}$ (not shown). Thus,  it becomes increasingly demanding to perform a quantum Fourier transform as the circuit size grows. 

To circumvent this issue, we now consider the approximate quantum Fourier transform, where small qubit rotations are neglected. To this end, we first analyze a controlled qubit rotation for different values of $k$. We can write the controlled rotation as 
\begin{equation}
CR_k = \ket{0}\!\bra{0}  \otimes 1 +  \ket{1}\!\bra{1}  \otimes  R_k,
\end{equation}
where $R_k$ is defined in Eq.~(\ref{eq:Rk}). A general two-qubit input state can be written as
\begin{equation}
\label{eq:twoqubitstate}
\ket{\Psi} =c_{00}\ket{00}+c_{01}\ket{01}+c_{10}\ket{10}+c_{11} \ket{11},
\end{equation}
where the complex coefficients $c_{ij}$ ensure that the state is normalized.  To understand the effects of a controlled rotation, we evaluate the distance between the initial state $\ket{\Psi}$ and the rotated state $\ket{\Psi'}=CR_k\ket{\Psi}$, defined as
\begin{equation}
\delta= 1-|\bra
{\Psi'}\!\Psi\rangle|^2
\label{eq:dist}
\end{equation} 
with explicit details in App.~\ref{appB}. In Fig.~\ref{fig:controlled_z_rotation_fidelity}, we show this distance as a function of the rotation angle, averaged over all possible two-qubit states. The figure shows us that to resolve rotations with $k > 4$, one needs a gate fidelity of at least $99\%$. Consequently, with lower gate fidelities, one might as well exclude rotations with $k\geq 5$. Leaving out those gates in a quantum Fourier transform then leads to a shallower quantum circuit.

In Fig.~\ref{fig:AQFT_fid}, we show the fidelity of the approximate quantum Fourier transform, which we define as
\begin{equation}
    f_l = |\braket{\Psi'_{0,l}}{\Psi_0}|^2,
\end{equation}
where  $\ket{\Psi_0}$ is the random input state, and 
\begin{equation}
 \ket{\Psi'_{0,l}} =  \mathrm{AQFT}_l^{-1}  \mathrm{QFT} \ket{\Psi_0}
\end{equation}
is the state after the quantum Fourier transform and the inverse approximate quantum Fourier transform with $k\leq l$ have been applied. For $l=2$, only controlled nearest-neighbor rotations are included, while $l=N$ corresponds to the exact quantum Fourier transform for a circuit with $N$ qubits. In Fig.~\ref{fig:AQFT_fid}, we use $N=20$ qubits, and we see that the fidelity does not improve much as $l$ is increased beyond $l=5$. Thus, in a quantum circuit with limited entanglement, it makes sense to truncate the rotations and apply the approximate quantum Fourier transform instead of the exact transform. In Fig.~\ref{fig:AQFT_fid}b, we show lines of constant fidelity as a function of the cut-off of the rotations. Again, we see that the fidelity of the approximate quantum Fourier transform is roughly constant if the cut-off $l$ is larger than five, noting that the fluctuations are due to different random initial states. 

Our results demonstrate how the fidelity of the transformation decreases as the entanglement is reduced.  In particular, the entanglement required to maintain the fidelity grows super-linearly with the number of qubits. Furthermore, we find that the fidelity of the approximate quantum Fourier transform is similar to the exact transformation if implemented with limited entanglement. A comparison left for future investigations is to explore if the same results would be obtained by applying the quantum Fourier transform as a single matrix product operator, as opposed to the gate-by-gate model followed here.

\subsection{Grover's search algorithm}

\begin{figure*}[t!]
    \centering
    \includegraphics[width=0.98\textwidth]{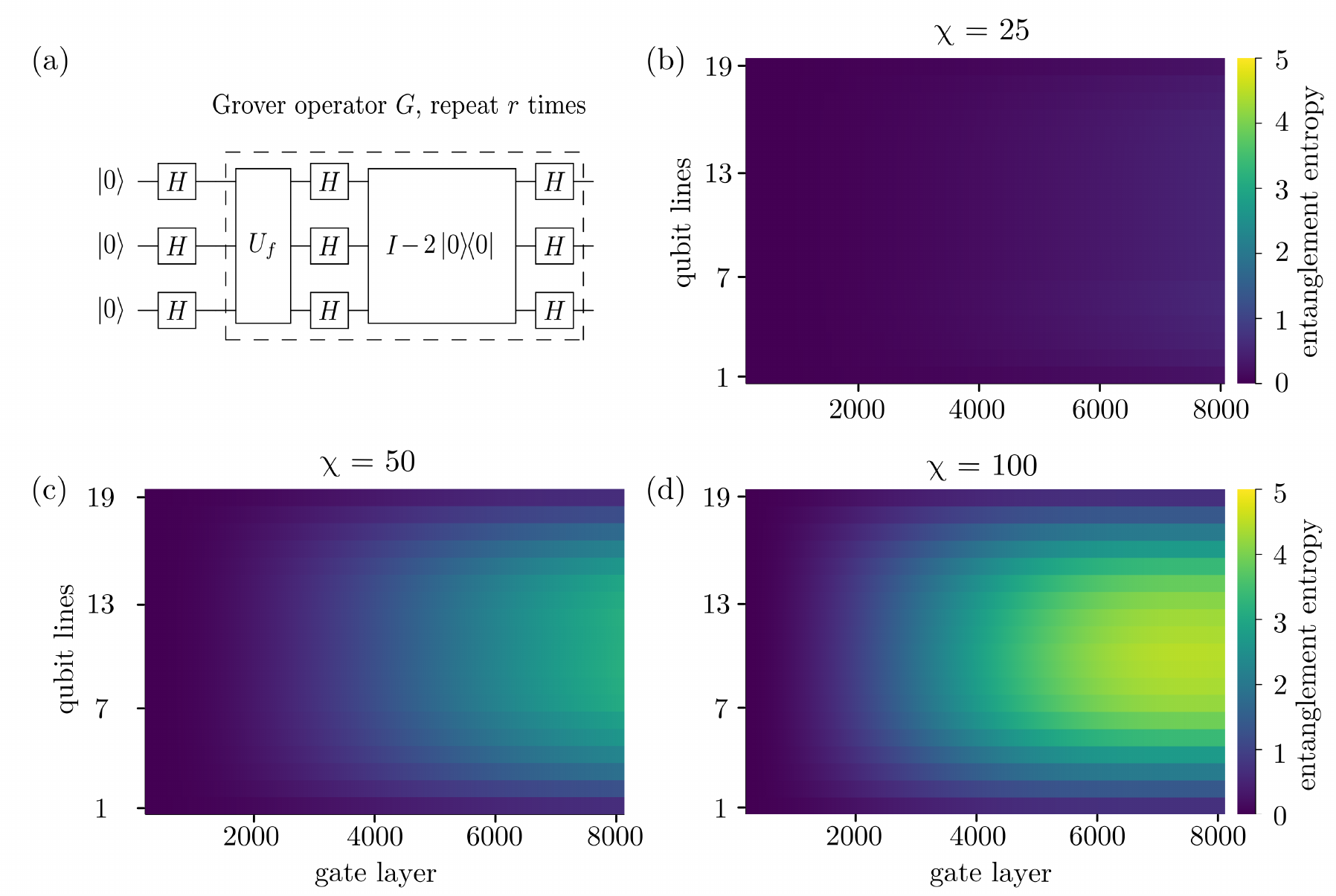}
    \caption{Entanglement in Grover's search algorithm. (a) Quantum circuit for Grover's algorithm. After the initial state is prepared in an equal superposition, the Grover oracle $U_f$ and the Grover diffuser are applied $r$ times to increase the probability of measuring the marked items in the database. We use the optimal number of iterations, which is approximately given by $r\simeq\sqrt{n/m}$ for $m$ marked items in a database with $n$ items. (b,c,d) Evolution of the entanglement entropy throughout the circuit with increasing bond dimension for $m=100$ marked items in a database with $n=2^{20}$ items encoded in $N=20$ qubits.}
    \label{fig:QC_Grover}
\end{figure*}

Grover's algorithm provides a quadratic speed-up for the task of searching an unsorted database~\cite{Grover1996}. In the following, we let $n$ denote the number of items in the database and $m$ the number of marked items that we are searching for. A circuit with $N$ qubits can store $n=2^N$ items, and the unsorted database is described by the state $\ket{s}=H^{\otimes N}\ket{0}$, which is an equal superposition of all items, and $H$ is the Hadamard gate. We now decompose the state of the database as
\begin{equation}
    \ket{s}=\cos(\alpha) \ket{\omega} + \sin(\alpha) \ket{\bar{\omega}}, 
\end{equation}
where we have introduced the angle $\alpha$, so that $\cos(\alpha) = \sqrt{m/n}$ and $\sin(\alpha) = \sqrt{(n-m)/n}$, and $\ket{\omega}$ is an even and normalized superposition of all marked items,
\begin{equation}
    \ket{\omega} = \frac{1}{\sqrt{m}}\sum_i^m \ket{\omega_i}.
\end{equation}
Similarly, all unmarked items are contained in $\ket{\bar{\omega}}$.  To find the marked items, the amplitude of $\ket{\omega}$ must be brought close to one. This amplitude amplification is achieved in two steps \cite{amplitude_amplification}. First, a Grover oracle $U_f$ flips the signs of all marked items
\begin{equation}
    U_f \ket{\psi} = 
    \begin{cases}
    &-\ket{\psi}, \quad \ket{\psi} \in \{ \ket{\omega_i} , i = 1, ..., m\}, \\
    &\quad \ket{\psi} \quad \text{otherwise}.
    \end{cases}
\end{equation}
Second, a Grover diffuser is applied, which implements a reflection about the equal superposition state as 
\begin{equation}
    G_\mathrm{diff}=1-2\ket{s}\! \bra{s} =H^{\otimes N}(1-2\ket{0} \! \bra{0} )H^{\otimes N}.
\end{equation}
The Grover operator $G=G_\mathrm{diff}U_f$ is now applied to the database $r$ times as shown in Fig.~\ref{fig:QC_Grover}a. This procedure increases the probability of measuring the marked items. The optimal number of iterations depends on the size of the database and the number of marked items and is roughly given by $\sqrt{n/m}$~\cite{Grover1996, nielsen_chuang_2010}. Thus, finding a single item requires about $\sqrt{n}$ iterations as compared to a classical algorithm, where $n/2$ iterations are needed. 

Here we implement the Grover oracle with prior knowledge of the element(s) to be identified, which allows us to implement the Grover oracle for any number of randomly drawn marked elements \cite{Figgatt2017}. This approach makes it possible to average over several draws and avoids any type of biases towards certain parts of the search space. More details on this issue are provided in App.~\ref{App_Grover}.

\begin{figure*}[t]
      \includegraphics[width=0.98\textwidth]{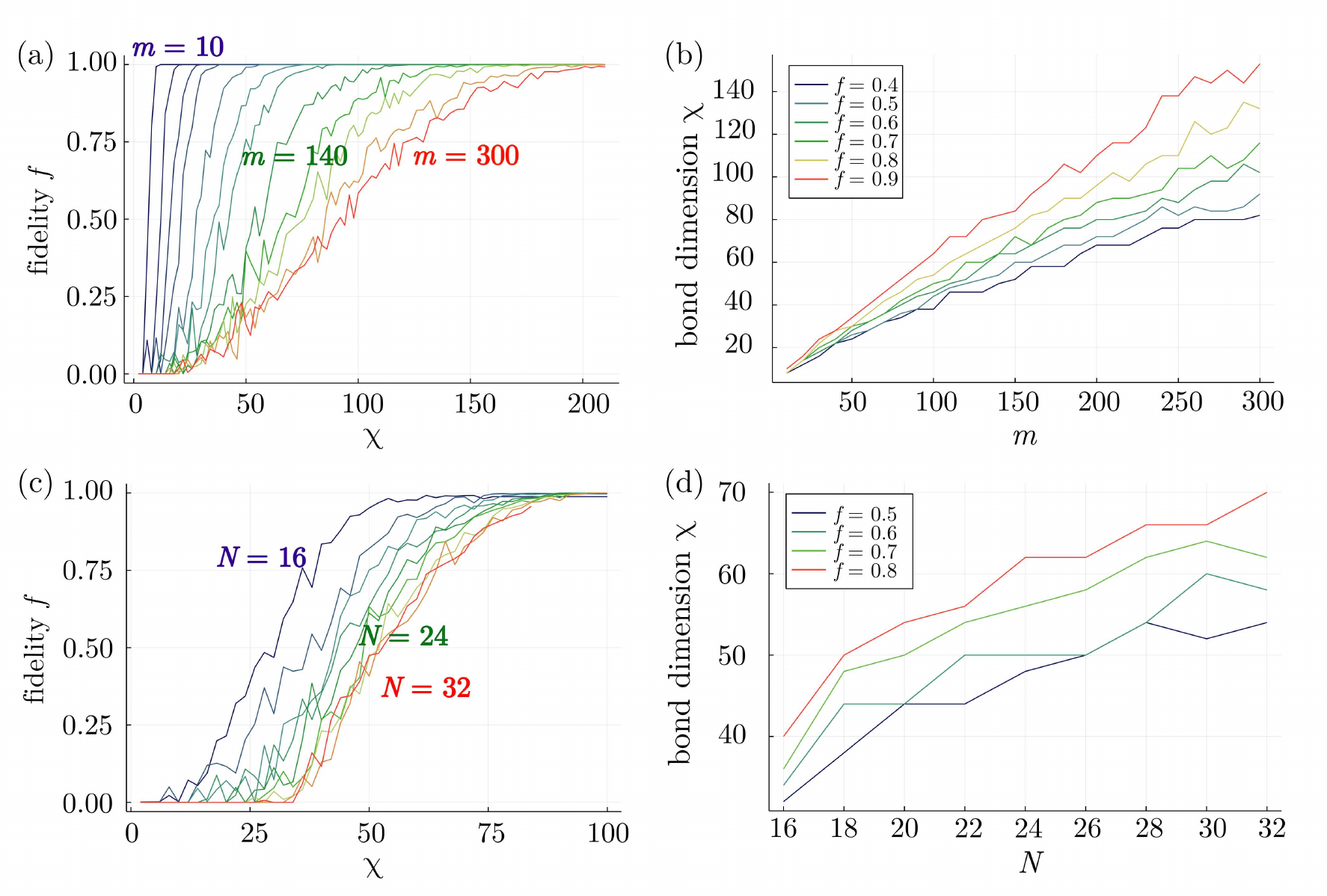}
    \caption{Fidelity of Grover's algorithm. (a) The fidelity as a function of the bond dimension for different numbers of marked items among a total of $n=2^{20}$ items encoded in $N=20$ qubits. (b) Lines of constant fidelity as functions of the number of marked elements. (c,d) Similar plots as in panels a and b, however, for different numbers of qubits with $m=100$ marked items.}
    \label{fig:Grover_fid}
\end{figure*}

We are now interested in the overlap between the marked items $\ket{\omega_i}$ and the state of the circuit after $r$ Grover iterations. Hence, we define the fidelity as
\begin{equation}
    f_r= \frac{1}{m}   \sum_i^m |\braket{\omega_i}{\Psi_r}|^2,
\end{equation}
where $\ket{\Psi_r}=G^r\ket{s}$. If the state $\ket{\Psi_r}$ is an even superposition of all marked elements, the fidelity is one. By contrast, it may be reduced below one, if the number $r$ of Grover rotations is not optimal, or if the algorithm is executed with limited entanglement. In the following, we use the optimal number $r$ of iterations,  such that the fidelity ideally would be close to one. We then examine how the fidelity is reduced, if we limit the entanglement.

Before discussing the fidelity, we first consider the entanglement in Grover's algorithm for different bond dimensions as shown in Figs.~\ref{fig:QC_Grover}b,c,d. Figure~\ref{fig:QC_Grover}d corresponds to the largest bond dimension, and one clearly sees how the entanglement in the circuit builds up as the number of Grover iterations approaches the optimal value. Thus, right before the state of the system is read out, the entanglement reaches its maximum, which is given by the entanglement in the superposition of all marked items. In the other panels, the bond dimension is lower, and the entanglement is partially lost. As such, we expect a significant reduction of the corresponding fidelity.

Figure~\ref{fig:Grover_fid} shows the fidelity of Grover's algorithm with limited entanglement. In Fig.~\ref{fig:Grover_fid}a, we show the fidelity as a function of the bond dimension for different numbers of marked items, which are drawn randomly at the beginning of each simulation~\cite{Figgatt2017, Mermin2007}. For different numbers of marked items, we average the results over $10$ random draws. The fluctuations in the results stem from two effects. First, since the marked items are generally different, the Grover operator will also be constructed in different ways. Therefore, it will be affected slightly differently by the truncation procedure to reduce the bond dimension. Second, the optimal number of rotations $r_{\text{opt}}$ is only defined up to an integer, and the final state in Grover's algorithm may not have a complete overlap with the marked items in the state $\ket{\omega}$. 

\begin{figure*}
    \centering
    \includegraphics[width=0.98\textwidth]{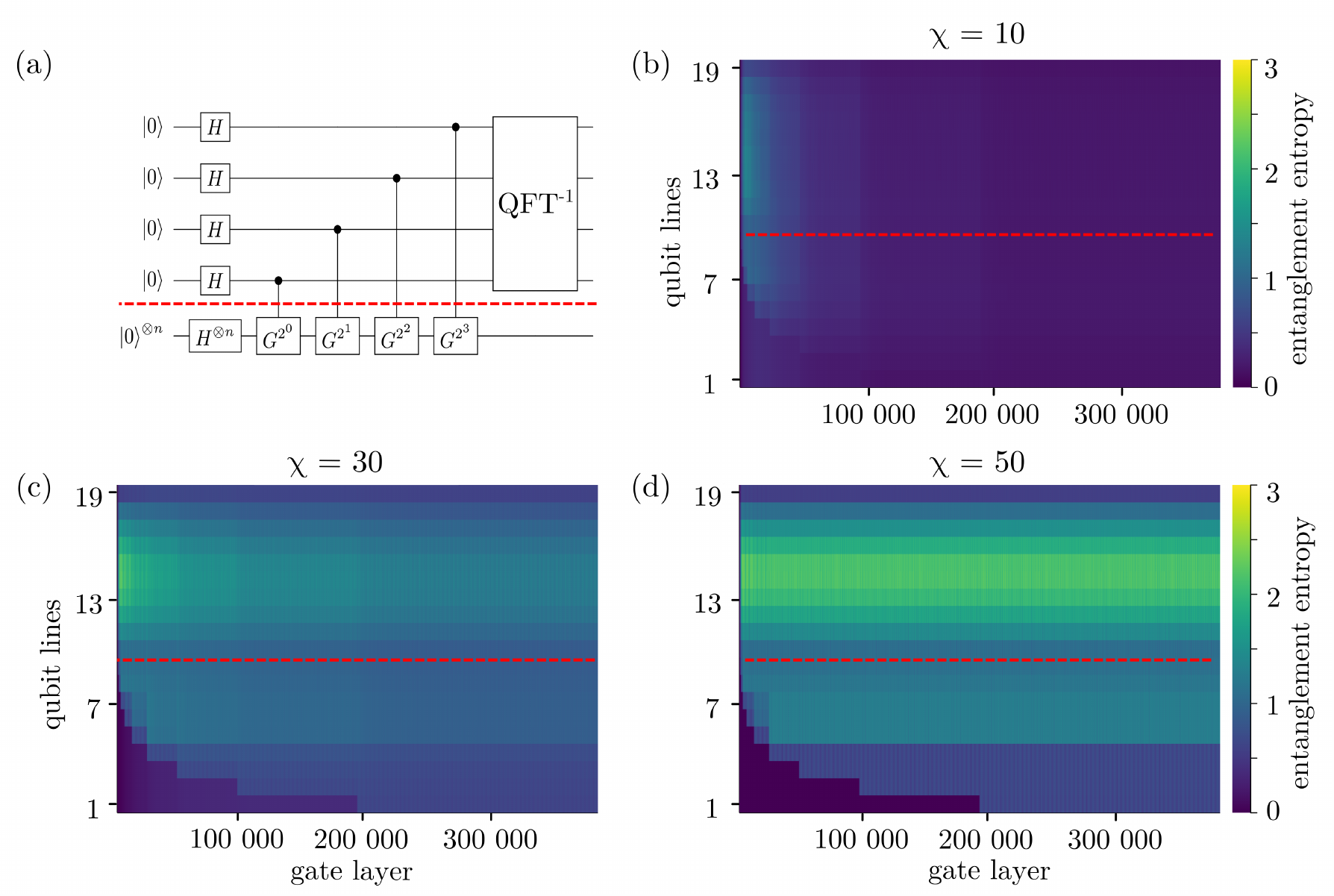}
    \caption{Entanglement in the quantum counting algorithm. (a) Quantum circuit for the quantum counting algorithm, which is equivalent to a quantum phase estimation applied to the Grover operator. The initial state is rotated into an equal superposition of all items in the database, which is the sum of two eigenstates of the Grover operator. (b,c,d) Entanglement entropy in the circuit for different bond dimensions with $N=20$ qubits and $m=30$ marked items of a total of $n=2^{10}=1024$ items.}
    \label{fig:QC_QC}
\end{figure*}

With a low number of marked items, the fidelity abruptly approaches one beyond a certain threshold, which can be remarkably low.  By contrast, for larger numbers of marked items, the fidelity increases more slowly with the bond dimension. Those results are of course consistent with the fact that choosing the bond dimension as $1+m$ implies a fidelity of one \cite{Stoudenmire2023}. This follows from the fact that a superposition of product states can be exactly represented as a matrix product state with a bond dimension given by the number of states. However, for all numbers of marked elements $m$ shown in Fig.~\ref{fig:Grover_fid}a, very high or even perfect fidelities are already obtained for bond dimensions lower than this upper bound.

Figure~\ref{fig:Grover_fid}b illustrates how the required bond dimension to achieve a certain fidelity grows approximately linearly with the number of marked items. This behavior can be contrasted with that of the quantum Fourier transform, where a super-linear growth of the bond dimension with increasing system size is observed. Figures~\ref{fig:Grover_fid}c,d show the dependence of the fidelity on the bond dimension for different numbers of qubits and a fixed number of marked items. Also, in this case, we observe that Grover's algorithm can be executed with high fidelity for low bond dimensions. We note that the required bond dimension to achieve a given fidelity depends roughly linearly on the number of qubits, compared to the super-polynomial dependence found in the case of the quantum Fourier transform in Fig.~\ref{fig:QFT_fid}b. This observation suggests that Grover's algorithm is more resilient to the loss of entanglement as compared to the quantum Fourier transform.

Furthermore, increasing the size of the search space requires only a modest increase in the bond dimension to achieve the same fidelity. Figure~\ref{fig:Grover_fid}d shows this explicitly for $100$ marked elements. Upon increasing the dimension of the search space from $2^{16} \simeq 6\times 10^4$ to $2^{32} \simeq 4\times 10^9$, an increase of only $20$-$30$ in the bond dimension is required to maintain a high fidelity. We finally note that, in the case of a single marked element, the entanglement entropy is bounded from above by $\ln(2)$, independently of the number of qubits~\cite{Stoudenmire2023}. Enlarging the search space here therefore merely leads to a linear increase in the computational cost through the increased system size, whereas the bond dimension can be kept constant.

Altogether, our results indicate that Grover's algorithm is more resilient to an increase in the total search space size than to an increase in the number of searched elements. In both cases, however, results with high fidelity have been obtained for comparatively low bond dimensions. Regarding the computational costs of our simulations, it is worth mentioning that a bond dimension of $\chi=100$ implies that only a tiny part of the Hilbert space is explored by the quantum algorithm, and, in the case of 32 qubits, this fraction can be estimated to be
\begin{equation}
    \text{fraction of Hilbert space} = \frac{2\times 32 \times \chi^2}{2^{32}} \simeq 10^{-4}.
\end{equation}
To simulate the algorithm with limited entanglement, one can further decrease the bond dimension, which again reduces the simulation cost. We emphasize that we only consider the reduction of the fidelity due to a loss of entanglement. However, it has been shown that the finite coherence time of the individual qubits leads to a  double-exponential decay of the fidelity with the qubit number \cite{Stoudenmire2023}. A comparison of this behavior with the dependence of the fidelity on the system size and the number of marked elements is left for future work.

To summarize, Grover's algorithm can be executed with high fidelity using a rather low bond dimension, in particular for a small number of marked items. As the number of marked items and the size of the database grow, the required bond dimension to achieve a given fidelity only increases linearly. For this reason, Grover's algorithm appears to be less sensitive to the loss of entanglement as compared to the quantum Fourier transform. 

\subsection{Quantum counting algorithm}

The problem of determining the number of marked items in a database, given a Grover operator, is solved by the quantum counting algorithm. It may also be applied to more general problems that can be recast as search problems~\cite{Brassard1998, nielsen_chuang_2010}. The quantum counting algorithm is an application of the quantum phase estimation algorithm \cite{Kitaev_QPE, nielsen_chuang_2010}, which can determine the eigenvalues of a unitary operator given the corresponding eigenstates. In the following, we seek the eigenvalues of the Grover operator. In the subspace spanned by $\ket{\omega}$ and $\ket{\bar{\omega}}$, the Grover operator can be represented by the matrix
\begin{equation}
    G= \begin{bmatrix}
    \cos(2\alpha) & -\sin(2\alpha)  \\
    \sin(2\alpha) & \cos(2\alpha) 
  \end{bmatrix}
\end{equation}
with eigenvalues $\lambda_\pm=e^{\pm 2\alpha i}$ and  corresponding eigenvectors $\ket{\pm}=(\pm\ket{\omega}+i\ket{\bar\omega})/\sqrt{2}$. The state of the database can be expressed as a superposition of these eigenstates,
\begin{equation}
    \ket{s}= (e^{-i\alpha}\ket{+}-e^{i\alpha}\ket{-})/\sqrt{2},
\end{equation}
which is the starting point for the quantum phase estimation algorithm to find the two eigenvalues. Once we have determined the phases of the eigenvalues, $\pm 2\alpha$, we obtain the number of marked items as~\cite{Brassard1998, nielsen_chuang_2010}
\begin{equation}
    m=  n \sin^2(\alpha),
\end{equation}
and the optimal number of iterations follows as 
\begin{equation}
    r_{\text{opt}} = \Bigg\lceil \frac{\pi}{4} \sqrt{\frac{n}{m}} \Bigg\rceil.
\end{equation}

Figure~\ref{fig:QC_QC}a shows the quantum counting algorithm. The quantum phase estimation part is implemented as described in Ref.~\cite{nielsen_chuang_2010}, and the Grover operators are constructed as in the previous section. This approach again has the advantage that we can initialize the Grover operator with a number of marked elements known to us and then compare the result of the algorithm to that number to verify its predictive capability at various levels of entanglement. The dashed red line in Fig.~\ref{fig:QC_QC}a separates an upper register, where the phases of the eigenvalues get encoded, and a lower one, where the Grover operator is applied. The phases are read out after the inverse quantum Fourier transform by measuring the first $N_\mathrm{read}$ qubits of the top register, which yields the phases with up to $N_\mathrm{read}$ binary digits. The output state has to be measured several times to construct a histogram of the outcomes, which will be peaked around the two phases. If the phases should be found with a probability of at least $1-\epsilon$, one has to construct the top register with
\begin{equation}
    N_\mathrm{aux} = \lceil \log_2 (2 + 1/2 \epsilon) \rceil
    \label{eq:succ_prob}
\end{equation}
auxiliary qubits, which are not measured at the end of the quantum phase estimation routine. The uncertainty in the number of marked items can be estimated as \begin{equation}
    \Delta m = \left(\sqrt{2m n} + n/2^{N_\mathrm{read}+1} \right)/ 2^{N_\mathrm{read}},
\end{equation}
which decreases with the number of qubits measured in the top register. By contrast, the success probability increases with the number of auxiliary qubits, which are not measured, according to Eq.~(\ref{eq:succ_prob}). For the following, we take $N_\mathrm{aux}=2$ corresponding to $\epsilon\simeq 0.09$. 

\begin{figure*}
\includegraphics[width=0.98\textwidth]{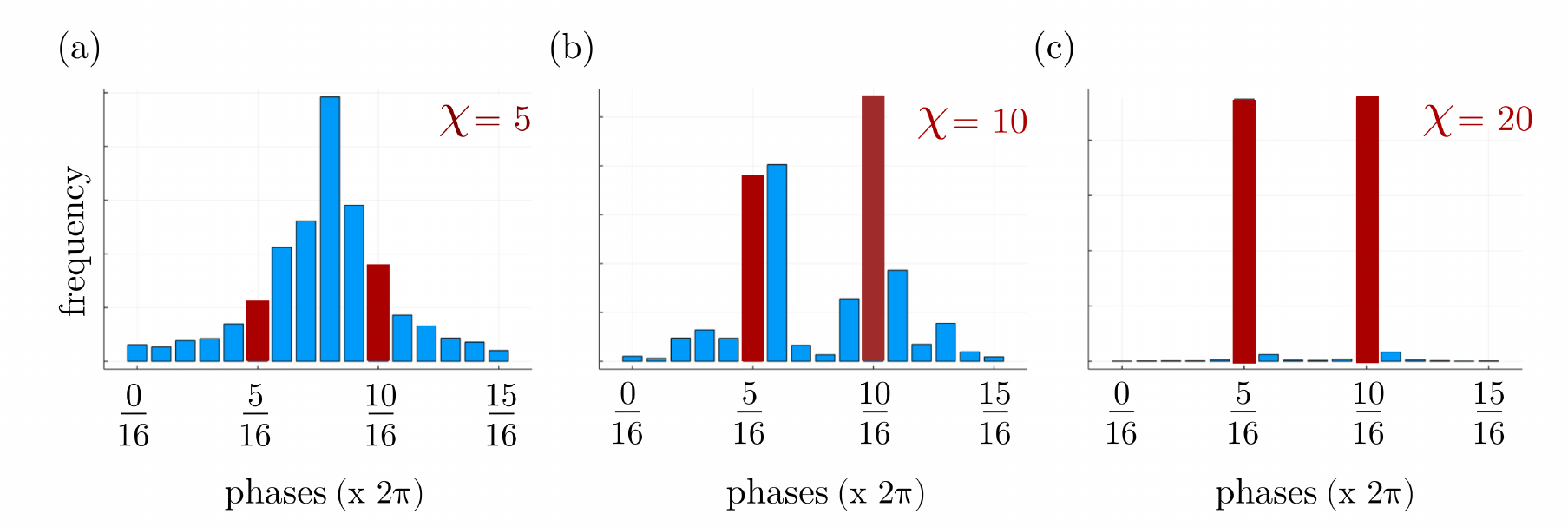}%
    \caption{Output of the quantum counting algorithm. (a,b,c) We consider a database with 30 out of 128 items marked. In the Fourier register, 4 out of 5 qubits are measured, so that the phases can be determined with up to 4 binary digits corresponding to a precision of $1/2^4\simeq 0.06$. The correct phases are marked in red. The panels correspond to different bond dimensions.}
    \label{fig:QCA_sampling}
\end{figure*}

\begin{figure*}[t]
    \includegraphics[width=0.98\textwidth]{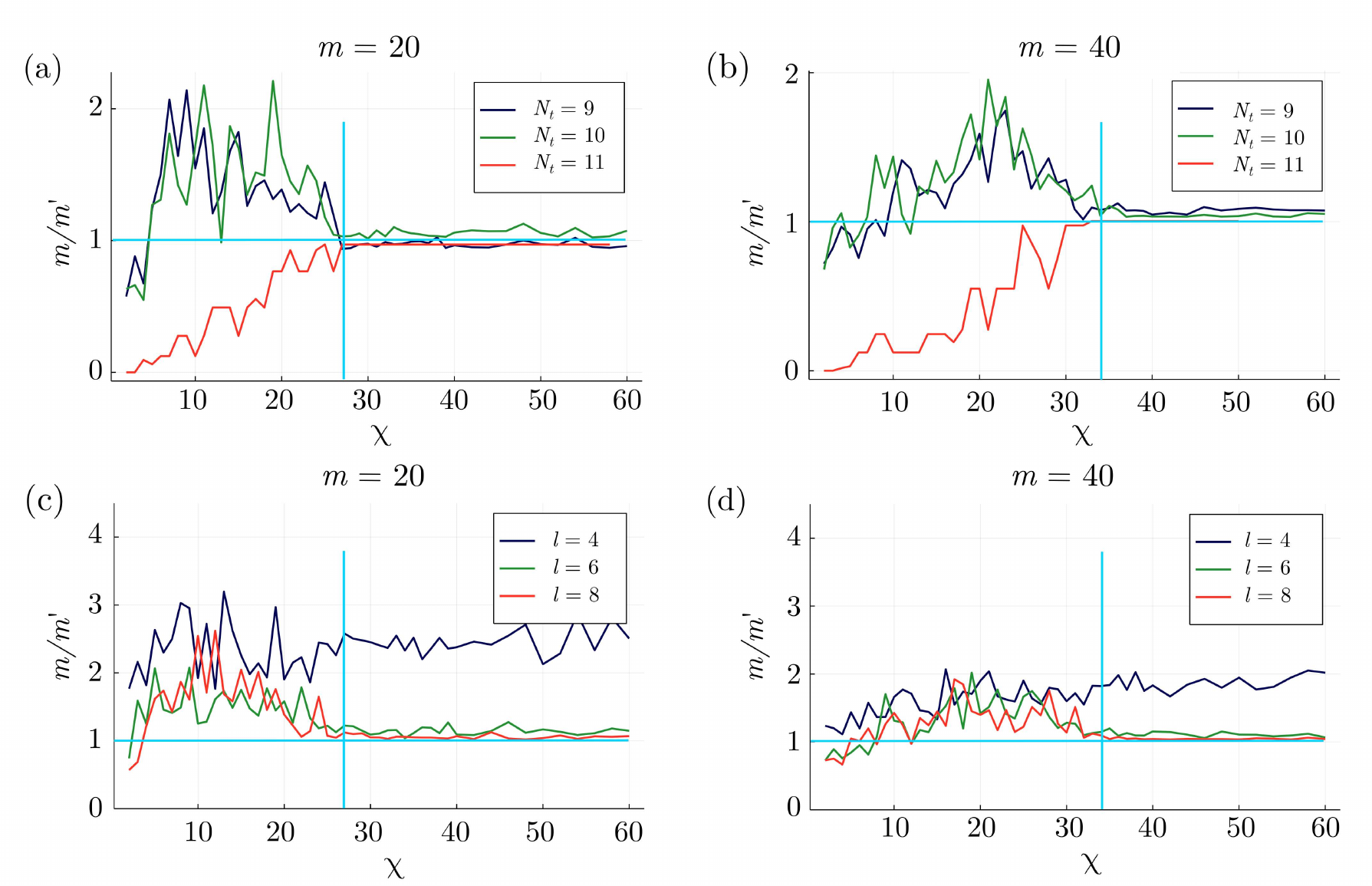}
    \caption{Accuracy of the quantum counting algorithm. The horizontal blue line marks the correct result of one, and the vertical blue line marks the bond dimension after which convergence is observed. (a,b) The number of marked items is determined by the quantum counting algorithm compared to the actual number as a function of the bond dimension. The lines correspond to different numbers of qubits in the quantum Fourier transform, while the Grover operator in each case is applied to 10 qubits. We consider $m=20$ and $m=40$ marked items in the two panels. (c,d) In these panels, we use the approximate quantum Fourier transform instead of the exact one, and we show results for different values of the truncation number $l$.}
    \label{fig:QC_MPS_num_sol}
\end{figure*}

In Figs.~\ref{fig:QC_QC}b,c,d, we show how the entanglement evolves throughout the circuit for different bond dimensions. Starting with the highest bond dimension in Fig.~\ref{fig:QC_QC}d, we see that the entanglement is concentrated in the upper register, which is conditioned on applications of the Grover operator in the lower register. In practice, this observation implies that the qubits with the highest fidelity should be used for the upper register since less entanglement is needed in the lower one. In Fig.~\ref{fig:QC_QC}b,c, the bond dimension is further reduced, leading to a loss of entanglement in the circuit with increasing circuit depth.

In the quantum counting algorithm, the state of the upper register is measured and converted into an approximation of the two phases, $\pm 2\alpha$. The algorithm is executed several times, and a histogram as shown in Fig.~\ref{fig:QCA_sampling} is generated. To determine the phases, we use a randomized method instead of simply selecting the two most probable values in the histogram. Instead, we randomly pick a value in the histogram based on its probability, and we then average over a large number of independent draws from the distribution. This method has the advantage of better dealing with statistical fluctuations for low bond dimensions, where the histogram might not be clearly peaked around the correct phases. With a high enough bond dimension, the method is effectively equivalent to simply selecting the two most probable phases.

The main outcome of the quantum counting algorithm is the number of marked items, and we now investigate how it depends on the bond dimension. To this end, we consider a $10$-qubit Grover register, corresponding to a database with $n=2^{10} = 1024$ items, together with $m=20$ or $40$ marked items. Between $6$ and $11$ qubits are used for the quantum Fourier transform with the last two of them not being measured as it increases the overall success probability. In Fig.~\ref{fig:QC_MPS_num_sol}, we show the number of marked items determined by the quantum counting algorithm over the actual number of marked items as a function of the bond dimension. In Figs.~\ref{fig:QC_MPS_num_sol}a,b, we consider $m=20$ and $m=40$ marked items in the database, respectively. The different lines correspond to varying numbers of qubits in the top register, which are used for the inverse quantum Fourier transform at the end of the algorithm. For large bond dimensions, the algorithm predicts the correct number of marked items, and the lines converge to one. This behavior depends only weakly on the number of qubits~$N_t$ in the quantum Fourier register. We note that the small fluctuations are due to the inherent quantum fluctuations and the probabilistic nature of the algorithm \cite{Brassard1998, nielsen_chuang_2010}. While our simulations in principle are deterministic, there are fluctuations because of simulated measurements at the end of the circuit.

In Figs.~\ref{fig:QC_MPS_num_sol}c,d, we show similar results, but this time using the approximate quantum Fourier transform instead of the full quantum Fourier transform. We use a quantum Fourier register with $N_t=10$ qubits and vary the truncation number of the approximate quantum Fourier transform between. With a truncation number of $l=6,8$, the results are similar to those obtained with the exact quantum Fourier transform. By contrast, for $l =4$, the algorithm does not identify the correct number of marked items because of the limited fidelity of the approximate quantum Fourier transform. Thus, we find that the quantum Fourier transform may be truncated to a gate fidelity of $99\%$ ($l=6$) and still yield the correct eigenvalues.

\section{Conclusions}
\label{conclusion}

We have simulated the execution of three quantum algorithms using matrix product states with a finite bond dimension to limit the available entanglement. Specifically, we have considered the quantum Fourier transform, Grover's search algorithm, and the quantum counting algorithm. In all three cases, we have found that the algorithms can be executed with a high fidelity even if the available entanglement is reduced. Our simulations also make it possible to map out the entanglement that is generated throughout the execution of each algorithm, allowing us to identify the regions of a circuit with the largest entanglement. For the quantum Fourier transform and Grover's algorithm, we have found that the fidelities decrease as the entanglement is reduced. However, for the quantum Fourier transform, the relationship is not linear in the system size. For the quantum counting algorithm, we have investigated how the prediction of the number of marked items depends on the loss of entanglement. Here we have found that the exact number can be extracted with relatively modest bond dimensions. This observation suggests that algorithms that are based on the quantum phase estimation can be executed even with relatively low entanglement. 

In the work presented here, we have focused on the reduction of  fidelities due to the loss of entanglement, which can be simulated using tensor-networks with a finite bond dimension. In an actual quantum computer, the fidelity can also be reduced due to other physical mechanisms, for instance, the single-qubit gates may be noisy. A possible inclusion of such error processes within the tensor-network formalism, we leave for future work. 

\section{Acknowledgments}
\label{Acknowledgments}
We acknowledge the computational resources provided by the Aalto Science-IT project as well as the financial support from InstituteQ, the Jane and Aatos Erkko Foundation, and the Academy of Finland through the Finnish Centre of Excellence in Quantum Technology (project numbers 352925 and 336810) and grant numbers 331342, 336243, and 308515.

\section{Data Availability}
\label{data_availability}
The Julia package used to produce the results is publicly available \cite{QSim_github} and contains user instructions as well as minimal examples. It is built on top of the \texttt{ITensors} package \cite{Itensors}, which provides the backend for the tensor-network calculations. Specific data generation scripts and data files are available upon request.

\appendix

\section{Controlled quantum gates}
\label{appendix:appA}

Here we provide a construction of single-qubit gates that are controlled by a number of other qubits. In particular, we show that they can be represented by a matrix product operator with a bond dimension of three. We note that bond dimensions of two have been achieved for similar matrix product operators, but we will not pursue this approach here~\cite{tensor_decomp_QC}.

To construct the matrix product operator, we consider a gate $U_j$ that acts on qubit $j$ and is controlled by $L$ other qubits that we label as $i_1, i_2, ..., i_L$. The gate is applied if all of the control qubits are in the state $\ket{1}$. The controlled operator can be written as
\begin{equation}
\begin{split}
     C_{i_1, i_2, ..., i_L}U_j &=  I_1 \otimes \cdots I_{i_1} \cdots I_j \cdots I_{i_L} \cdots \otimes I_N \\
    - & I_1 \otimes \cdots \ket{1}\!\bra{1}_{i_1}\cdots I_j \cdots \ket{1}\bra{1}_{i_L}  \cdots \otimes I_N \\
    + & I_1 \otimes \cdots \ket{1}\!\bra{1}_{i_1} \cdots U_j \cdots \ket{1}\!\bra{1}_{i_L}  \cdots \otimes I_N, \\
\end{split}
\end{equation}
where the first line expresses that the operator generally acts trivially on the qubits. Only if all control qubits are in the state $\ket{1}$, the gate $U_j$ is applied to the target qubit instead of $I_j$ as expressed by the two other lines.

Similar to the conversion of a locally interacting Hamiltonian into the form of a matrix product operator \cite{Hubig2017}, we can now find the abstract tensors in matrix form, which can be multiplied to obtain the three terms in the expression above. For the sake of concreteness, we show the construction for the five-qubit operator $C_{1, 4, 5}\,X_3$,
\begin{equation}
    \underset{\text{control}}{\begin{bmatrix}
    I_1 \\
    -P_1 \\
    P_1 \\
    \end{bmatrix}^{T}}
    \underset{\text{spectator}}{\begin{bmatrix}
    I_2 &  & \\
    & I_2 & \\
    &  & I_2 \\
    \end{bmatrix}}
    \underset{\text{target}}{\begin{bmatrix}
    I_3 &  & \\
    & I_3 & \\
    &  & X_3 \\
    \end{bmatrix}}
    \underset{\text{control}}{\begin{bmatrix}
    I_4 &  & \\
    & P_1 & \\
    &  & P_1 \\
    \end{bmatrix}}
    \underset{\text{control}}{\begin{bmatrix}
    I_5 \\
    P_1 \\
    P_1 \\
    \end{bmatrix}},
\end{equation}
where $P_1 =\ket{1}\!\bra{1}$ is a projector, the second qubit is just a spectator qubit, and the multiplication of individual entries should be understood as tensor products.

Evaluating these matrix products indeed yields the correct form of the operator. A generalization to more control qubits as well as qubit registers of different sizes is straightforward by inserting the corresponding matrices for the control and the spectator qubits. Furthermore, this construction can  be extended to several controlled single-qubit operators by replacing a spectator tensor with an additional target tensor as
\begin{equation}
    \underset{\text{spectator}}{\begin{bmatrix}
    I_i &  & \\
    & I_i & \\
    &  & I_i \\
    \end{bmatrix}} \rightarrow
    \underset{\text{target}}{\begin{bmatrix}
    I_i &  & \\
    & I_i & \\
    &  & U_i \\
    \end{bmatrix}}.
\end{equation}
With this approach, one can construct any controlled gate as a matrix product operator of bond dimension three (since the matrices above are of rank three). Hence, we can implement all the operators that we need for quantum computations, namely single-qubit gates and controlled gates, with relatively low computational costs.

\section{Sampling from a matrix product state}
\label{appendix:app_sampling}

Here we briefly summarize the algorithm to draw a sample from the probability distribution represented by a matrix product state~\cite{Ferris2012, Itensors}. The probability $p(\boldsymbol{\sigma})$ to draw a sample $\{\sigma_1, \sigma_2, ..., \sigma_N \}$ can be decomposed into a product of conditional probabilities:
\begin{equation}
    p(\boldsymbol{\sigma}) = \prod_{i=1}^N p(\sigma_i | \sigma_1 ... \sigma_{i-1}).
\end{equation}
Assuming a state $\ket{\Psi}$ in right-canonical form, the conditional probabilities $p(\sigma_i | \sigma_1 ... \sigma_{i-1})$ can be constructed sequentially with a right-sweep through the matrix product state to obtain the local reduced density matrices. We start by constructing the reduced density matrix 
\begin{equation}
    \left[ \rho_1 \right]^{\sigma_1, \sigma_1'}  = \sum_{ \chi_{1}} [A_1]_{1 \chi_1}^{\sigma_1} [A_1^\dagger]_{1 \chi_1}^{\sigma_1'},
    \label{eq:reduced_rho}
\end{equation}
of the first qubit. This contraction is the only one that needs to be performed, guaranteed by the right-canonicity of the state $\ket{\Psi}$. The sample $\{\sigma_1 \}$ is then generated by drawing a random number according to the probability distribution defined by the diagonal elements of $\rho_1$. In the next step, to draw a sample according to the conditional probability $p(\sigma_2 | \sigma_1)$, we project the previous measurement result into the tensor $A_2$ by letting
\begin{equation}
    [A_2]_{\chi_1 \chi_2}^{\sigma_2} \rightarrow \sum_{\chi_1} [A_1]_{1 \chi_1}^{\sigma_1} [A_2]_{\chi_1 \chi_2}^{\sigma_2} \equiv [B_2(\sigma_1)]_{1 \chi_2}^{\sigma_2}
\end{equation}
and by re-weighting the matrix product state by $1/\sqrt{p(\sigma_1)}$. The reduced density matrix now reads
\begin{equation}
    \left[ \rho_2 (\sigma_1) \right]^{\sigma_2, \sigma_2'}  = \sum_{ \chi_{2}} [B_2]_{1 \chi_2}^{\sigma_2} [B_2^\dagger]_{1 \chi_2}^{\sigma_2'},
    \label{eq:reduced_rho}
\end{equation}
and the next element $\{\sigma_2 \}$ in the sample is then by construction drawn according to the resulting probability distribution $p(\sigma_2 | \sigma_1)$. This procedure is iterated until a full right-sweep has been completed.

\section{General two-qubit state}~\label{appB}

Here we provide additional details of our calculations in Fig.~\ref{fig:controlled_z_rotation_fidelity}.
By imposing that the two-qubit state in Eq.~(\ref{eq:twoqubitstate}) is normalized  and invariant under global phase shifts, we may write the expansion coefficients as
\begin{equation}
\begin{split}
c_{00}&=\cos{ \phi_1},\\ 
c_{01}&=\sin{ \phi_1} \cos{ \phi_2}  e^{i  \Phi_1},\\
c_{10}&=\sin{ \phi_1} \sin{\phi_2} \cos{ \phi_3}  e^{i  \Phi_2}, \\
c_{11}&=\sin{\phi_1} \sin{ \phi_2} \sin{ \phi_3} e^{i  \Phi_3}
\end{split}
\end{equation}
where the parameters $\phi_1, \phi_2$ run in the interval $[0, \pi]$, and $\phi_3, \Phi_1, \Phi_2, \Phi_3$ in the interval $[0, 2\pi)$. This parametrization can be obtained by representing each coefficient $c_{ij}$ in polar form as $r_{ij}\exp(i \phi_{ij})$, with both $r_{ij}$ and $\phi_{ij}$ being real numbers. Since $\sum_{ij}|r_{ij}|^2 = 1$, we may represent the $r$-coordinates as spherical coordinates on a $4$-sphere, which yields the explicit dependence on the sine- and cosine functions. Factoring out a global phase factor, e.g. $\exp(i \phi_{00})$, and redefining the other phase angles accordingly, we arrive at the parametrization above.

For the overlap between any initial state $\ket{\Psi}$ and the rotated state $\ket{\Psi'}=CR_k\ket{\Psi}$, we find

\begin{equation}
\begin{split}
       |\bra{\Psi'}\!\Psi\rangle|^2 =\Bigg[&\cos^2(\phi_1) + \sin^2(\phi_1)\cos^2(\phi_2)  \\
       &+ \sin^2(\phi_1)\sin^2(\phi_2)\cos^2(\phi_3) \\
       &+ \sin^2(\phi_1)\sin^2(\phi_2)\sin^2(\phi_3)\cos(\frac{2\pi}{2^k})\Bigg]^2\\       &+\Bigg[\sin^2(\phi_1)\sin^2(\phi_2)\sin^2(\phi_3)\sin(\frac{2\pi}{2^k}) \Bigg]^2, 
\end{split}
\end{equation}
which turns out to be independent of the phases $\Phi_1, \Phi_2$ and $\Phi_3$. Using this expression, we find the maximum distance and the mean and median values shown in Fig.~\ref{fig:controlled_z_rotation_fidelity}.

\section{Construction of Grover oracles}
\label{App_Grover}

\begin{figure}
    \includegraphics[width=\columnwidth]{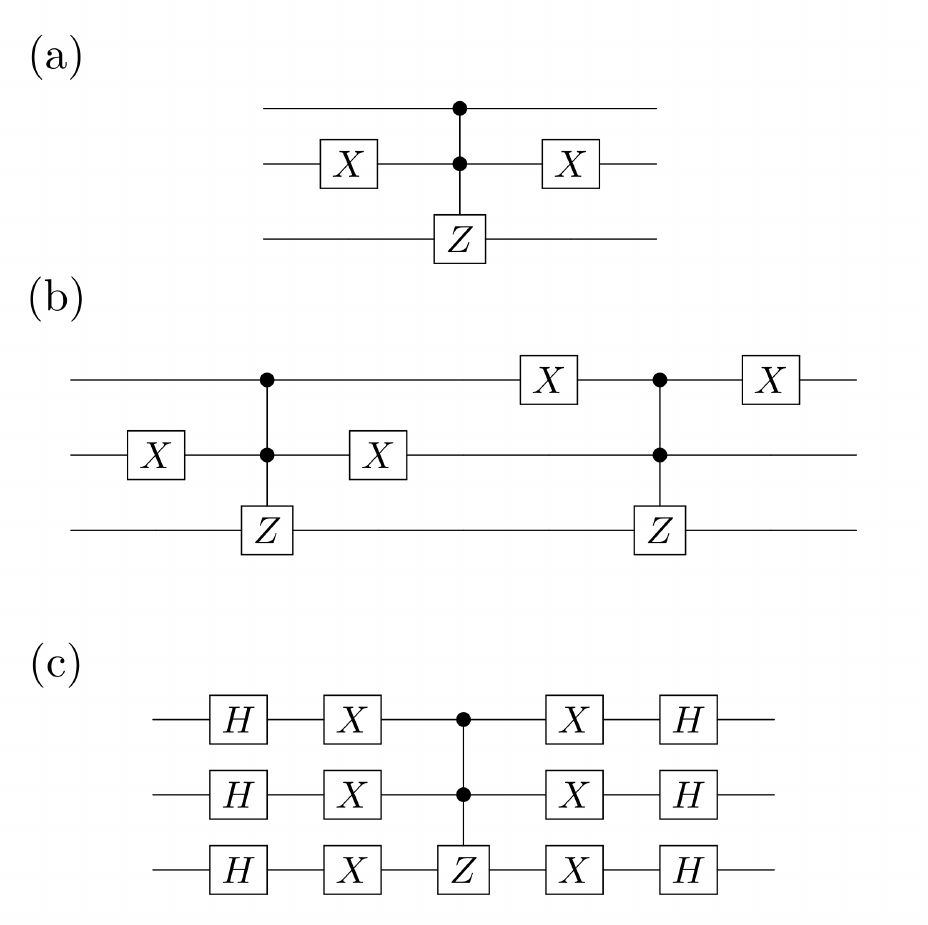}%
    \centering
    \caption{Explicit construction of Grover oracles and diffuser for a $3$-qubit system. (a) Grover oracle marking the state $\ket{101}$. (b) Grover oracles marking the states $\ket{101}$ and $\ket{011}$. (c) Grover diffuser.}
    \label{fig:Grover_oracle}
\end{figure}

The center piece of Grover's algorithm is the construction of the Grover oracle $U_f$, which is able to identify the solutions to a given search problem. Here, we investigate the fidelity of Grover's algorithm and its resource requirements as a function of the number of marked elements and the system size. We have therefore chosen the method of first randomly selecting $m$ marked elements and then constructing the corresponding Grover oracle $U_f$, which would mark those states as realized experimentally in Ref.~\cite{Figgatt2017}. An example of this construction for $3$ qubits is shown in Fig.~\ref{fig:Grover_oracle}a. Here, the basis state to be identified is mapped to $\ket{111}$ through a suitable sequence of \texttt{X}-gates. Afterwards, the controlled \texttt{Z}-gate produces the required phase flip. This construction immediately extends to several marked elements as in Fig.~\ref{fig:Grover_oracle}b. The Grover diffusor can be implemented as shown in Fig.~\ref{fig:Grover_oracle}c.

This implementation can be interpreted as follows: Let a database be constructed from $N$ pairs of elements, each in one-to-one correspondence to each other. An example of this would be a phone book with a mapping between names and phone numbers. Basis states describing this database could then be enumerated as the tensor product 
\begin{equation}
    \ket{\text{key}_i} \otimes \ket{\text{value}_i}, \quad i = 1, ..., N.
\end{equation}
Typically, the keys in such a scenario would be considered as known and  ordered, e.~g.~alphabetically. The collection of values, however, can be of an unstructured nature. If one wanted to find the key corresponding to a given value, one could therefore run Grover's algorithm with the oracle construction on the value-register, and identify the corresponding key. The input for the oracle would be known at all stages of algorithm, yet the search problem remains non-trivial. We would like to emphasize that this construction is especially well-suited for matrix product operator simulations of multiply-controlled quantum gates, as increasing the number of control qubits does not increase the bond dimension of the matrix product operator. It  is therefore straightforward to scale this construction to any number of qubits.

This method of implementing a Grover oracle is to be contrasted with the oracle corresponding to a search problem, where the solutions are not known~\cite{quantum_search_hash_functions}. Examples of this are satisfiability problems, where a certain bit string satisfying certain logic conditions is sought. Here, the oracle merely needs to recognize the solution, which is possible by a direct encoding of the logic conditions. 

\bibliography{biblio}

\end{document}